\documentclass[osajnl,showpacs,amsmath,amssymb,english,floatfix]{revtex4}
\usepackage{graphicx}
\usepackage{amsmath}\usepackage{amsbsy}
\usepackage{dcolumn}
\usepackage{bm}
\usepackage[T1]{fontenc}
\usepackage[latin1]{inputenc}

\providecommand{\tabularnewline}{\\}

\usepackage{ae}

\usepackage{babel}

\begin{document}

\title{Low-loss resonant modes in deterministically aperiodic nanopillar
waveguides}

\author{Sergei V. Zhukovsky, Dmirty N. Chigrin, Johann Kroha}

\affiliation{Physikalisches Institut, Universit\"at Bonn,\\
Nussallee 12, 53115 Bonn, Germany}

\email{sergei@th.physik.uni-bonn.de}

\begin{abstract}
Quasiperiodic Fibonacci-like and fractal Cantor-like single- and multiple-row nanopillar waveguides are investigated theoretically employing the finite difference time domain (FDTD) method. It is shown that resonant modes of the Fibonacci and Cantor waveguides can have a Q-factor comparable with that of a point-defect resonator embedded in a periodic nanopillar waveguide, while the radiation is preferably emitted into the waveguide direction, thus improving coupling to an unstructured dielectric waveguide located along the structure axis. This is especially so when the dielectric waveguide introduces a small perturbation in the aperiodic structure, breaking the structure symmetry while staying well apart from the main localization area of the resonant mode. The high Q-factor and increased coupling with external dielectric waveguide suggest using the proposed deterministically aperiodic nanopillar waveguides in photonic integrated circuits.
\end{abstract}

\maketitle

\section{Introduction\label{sec:Introduction}}

In the recent decade, defects in two-dimensional (2D) photonic crystals
have been widely considered for use as photonic microcavities 
\cite{jmwBook,sakodaBook}. A substantial amount of work has been devoted to optimizing the Q-factor of the resonant mode while maintaining the small mode volume (see \cite{Recipe} and references therein). Since a 2D photonic crystal does not possess a complete 3D photonic band gap, numerous efforts were dedicated to determining the influence of cladding or substrate on the resonator quality  \cite{Deloc1,Deloc2}. There have been efforts to utilize resonant modes in a 2D system for microlasers \cite{Deloc2,IEICE} and to employ the resonant mode formalism in random lasing studies \cite{cao05,Deloc3}.

As it has recently been discussed, periodic one-dimensional (1D)
arrangement of dielectric nanopillars (\emph{nanopillar waveguides}) can support
guided modes \cite{Fan95,AVL,AVL2}. Such a dielectric structured waveguides possess good light confinement along with the strongly modified dispersion. The current state-of-the-art fabrication technologies makes practical realization of the nanopillar structures feasible. To ensure the vertical light confinement and low losses, the pillar-based dielectric structures can be realized both in sandwich-like geometry \cite{pillar04,pillar05} and in membrane-like geometry \cite{ParkLee}. Along with known studies of lasing in single nanowires and nanowires arrays \cite{Lasing}, it is suggested that nanopillar geometry, and in particular 1D nanopillar arrangements, can prove a good alternative to already known 1D and 2D microstructured waveguides and integrated optical devices based on them. 

\begin{figure}[h]
\includegraphics[clip,width=\columnwidth]{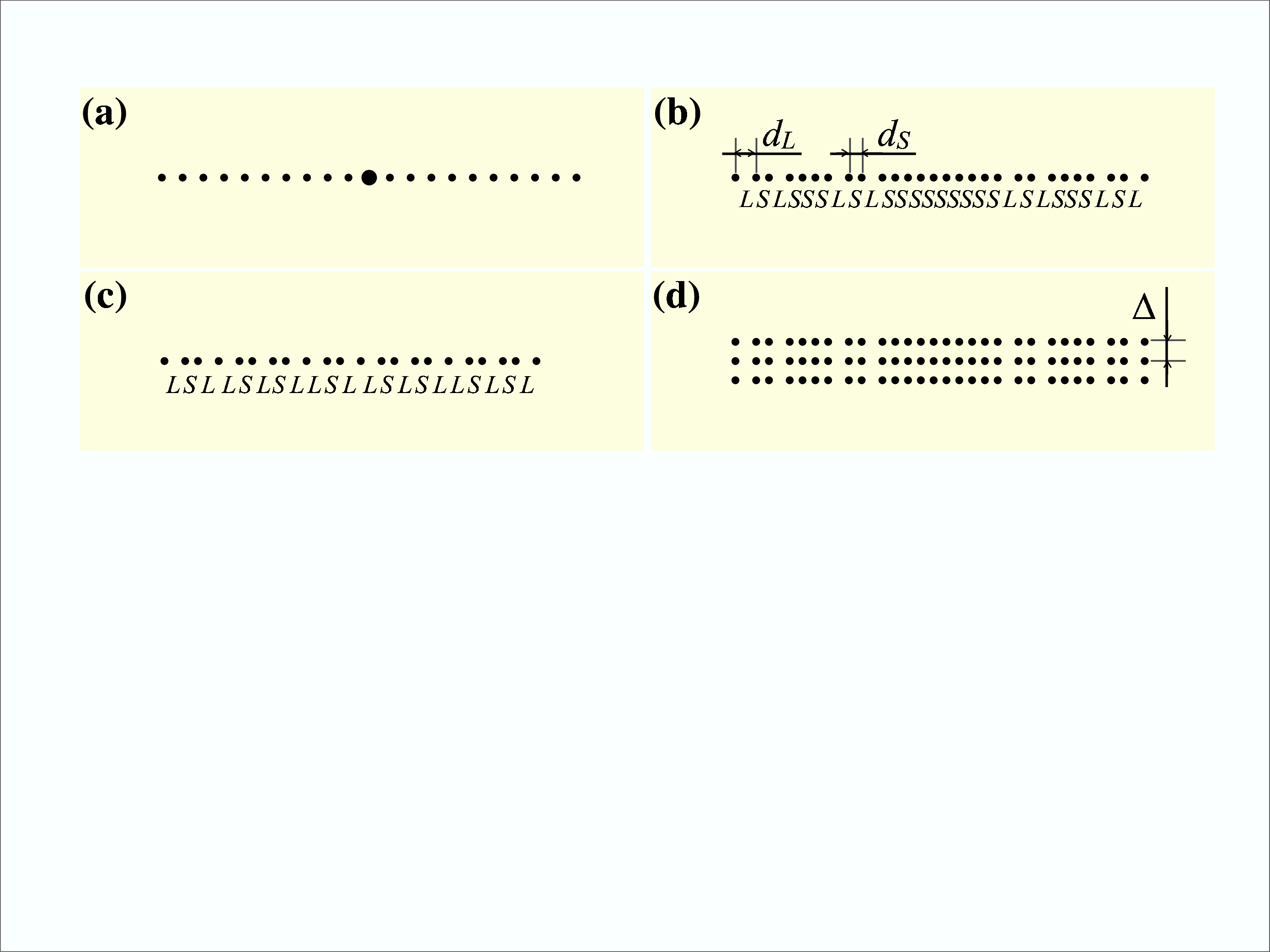}
\caption{Top view of a point-defect periodic reference waveguide \textbf{(a)},
a third-generation Cantor nanopillar waveguide \textbf{(b)},
a 7-stage Fibonacci waveguide \textbf{(c)},  
and a three-row Cantor waveguide
with row spacing $\Delta=0.75a$ \textbf{(d)}. 
Black circles correspond to cylinders with $\varepsilon=13.0$, 
extending infinitely in the third dimension and located in air. 
For the FDTD calculations a spatial grid with 16 mesh points per
unit length $a$ was used. The whole area of computation 
is a $7a\times22a$ rectangle surrounded by PML boundaries, 
see section \ref{sec:setup} for details.
\label{fig:structure}}
\end{figure}

Most applications involve introducing a defect
into a waveguide-like structure, e.g.~for use as a microresonator, coupled to
a 1D waveguide as an input or output terminal. 
Usually, the defect is considered to be point-like,
created by altering the properties of one or several adjacent
nanopillars (see, e.g., Fig.~\ref{fig:structure}a) \cite{Fan95,Johnson}. 
However, due to the absence of a complete band gap, 
the breaking of translational symmetry caused by the defect inevitably results
in radiation losses of the microresonator mode by coupling to the 
surrounding electromagnetic continuum. The losses should be much 
stronger in the case of 1D nanopillar waveguides than in the more studied case of 2D periodic structures. This raises the need for optimizing the Q-factor of the resonator in 1D nanopillar waveguides. If one were to couple the resonant mode efficiently to an external
unstructured dielectric waveguide, located along the nanopillar waveguide axis, and if the amount of energy escaping through this dielectric waveguide were greater than
that leaving the resonator elsewhere, this would prove useful in many design aspects of nanosized optical components, for instance microlaser resonators. 

There have been some proposals to decrease the losses based on either
mode delocalization \cite{Deloc1} or on the effect of multipole cancellation
\cite{Johnson}. However, a delocalized mode typically suffers from a decrease of the Q-factor. On the other hand, the spatial radiation loss profile of a mode described in \cite{Johnson}
has a nodal line along the waveguide axis, which means poor coupling to any components coaxial with the waveguide.

Other than by means of a point defect, a resonant system can also be created by changing the periodic arrangement of nanopillars into non-periodic. There is a wide variety of possibilities for creating such a {\em deterministically aperiodic} structure. The most studies cases of such structures are quasiperiodic (e.g. Fibonacci-like) \cite{Kohmoto} and fractal (e.g. Cantor-like) \cite{PRE} structures. The examples are shown in Fig.~\ref{fig:structure}b,c.
In a 1D multilayer model, such deterministically aperiodic structures are known to support eigenmodes which are highly localized yet with the localization character different from exponential \cite{Kohmoto}, as well as with localization patterns extending over several similar defects throughout the structure \cite{PRE}. In addition, several of such non-periodic waveguides can easily be aligned next to each other to form a multi-row nanopillar waveguide 
(see Fig.~\ref{fig:structure}d) \cite{AVL,AVL2}. It causes waveguide modes to split, as it always happens when several resonant systems are coupled. What is more, each of these split modes
can be selectively excited if the excitation pattern resembles the profile of the corresponding mode \cite{AVL}.

In this paper, we investigate the resonant modes in single- and multi-row nanopillar waveguides constructed according to two known deterministically aperiodic sequences -- a quasiperiodic Fibonacci \cite{Primer} and a middle-third Cantor fractal sequence \cite{EPL} -- and compare their
properties to those of the resonant modes of a point-defect reference system \cite{Johnson}. We show that, although the resonant modes of a point defect generally exhibit a higher Q-factor
and always have a smaller mode volume, in a 1D periodic nanopillar arrangement the coupling to an external terminal (a semi-infinite rectangular unstructured waveguide placed coaxially with nanopillar waveguide) is poor, unless the terminal is brought very close to the defect. On the contrary,
both in a Fibonacci and in a Cantor waveguide there is a good coupling to the terminal while maintaining a high Q-factor. The coupling increases even further when the terminal breaks the symmetry of the Fibonacci or Cantor structure, yet stays well apart from the mode's main localization region. Both the Q-factor and the coupling are found to improve even further if multiple-row Cantor nanopillar waveguides are used.

The paper is organized as follows. In Section~\ref{sec:Q-factor} we describe the construction of quasiperiodic and fractal nanopillar waveguides. The analysis of the resonant modes Q-factors is presented for single- and multiple-row deterministic aperiodic waveguides. In Section~\ref{sec:Coupling} we discuss the coupling of resonant modes with an external unstructured dielectric waveguide and show a distinct advantage in both Cantor and Fibonacci cases over a reference point-defect system. In Section~\ref{sec:Multiple} the multiple-row Cantor waveguides are studied. Finally, Section~\ref{sec:Conclusion} summarizes the paper.

\section{Resonant modes in aperiodic waveguides\label{sec:Q-factor}}

\subsection{Generation of aperiodic structures and computational method
\label{sec:setup}}

As a system under study, we have chosen a non-periodic 1D array of
cylindrical nanopillars of equal diameter, the distances between adjacent
pillars given by Fibonacci and Cantor sequences, respectively. If we denote~$S$
and~$L$ for short and long distance ($d_{S}$~and~$d_{L}$), respectively,
the Fibonacci sequence is constructed by the inflation rule,
\begin{equation}
L\to LS,\quad S\to L,\label{eq:fib_subst}
\end{equation}
and reads,
\begin{eqnarray}
L&\to& LS\to LSL\to LSL\, LS  \\
 &\to& LSLLS\, LSL\to LSLLSLSL\, LSLLS\to\cdots \ . \nonumber
\label{eq:fib_example}
\end{eqnarray}
The Cantor sequence is created by the inflation rule
\begin{equation}
L\to LSL,\quad S\to SSS.\label{eq:cantor_subst}
\end{equation}
and unfolds in the following
self-similar fashion, which represents a series of middle third Cantor
prefractals
\begin{eqnarray}
L&\to& LSL\to LSL\, SSS\, LSL \\
 &\to& LSLSSSLSL\, SSSSSSSSS\, LSLSSSLSL\to\cdots.
\nonumber
\label{eq:cantor_example}
\end{eqnarray}
For the Fibonacci sequence, it is remarkable that the distance from
the first pillar (which has number~$0$) to the $n$th pillar can be expressed
via the simple, explicit relation \cite{Primer},
\begin{equation}
d(n)=d_{S}n+\left(d_{L}-d_{S}\right)\left\lfloor \frac{n-1}{\tau}\right\rfloor ,\label{eq:fib_analytic}
\end{equation}
$\tau=(1+\sqrt{5})/2$ being the golden mean. We have found that a
similar formula for a Cantor sequence can be derived, albeit considerably
more complicated,
\begin{equation}
d(n)=d_{S}n+\left(d_{L}-d_{S}\right)\sum_{j=1}^{n}\prod_{k=1}^{\left\lceil N\right\rceil }\left(1-2\left\{ \frac{1}{2}\left\lfloor \frac{j-1}{3^{k}}\right\rfloor \right\} \right).\label{eq:cantor_analytic}
\end{equation}
In Eqs.~(\ref{eq:fib_analytic}--\ref{eq:cantor_analytic}), the
expressions~$\left\lfloor x\right\rfloor $,~$\left\lceil x\right\rceil $,
and~$\left\{ x\right\} $ denote floor integer, ceiling integer,
and fractional parts of~$x$, respectively, whereas $N=\log_{3}n$.
For the realizations of our aperiodic structures we have chosen 
for the radius of a nanopillar $r=0.15a$ and for the short and long spacings
between adjacent nanopillars $d_{S}=0.5a$ and $d_{L}=0.81a$, respectively. The pillars have a dielectric constant $\varepsilon =13$ and are placed in air ($\varepsilon =1$). 
Fig.~\ref{fig:structure} shows a ``top view'' of a 
3rd-generation Cantor structure consisting of $28$ nanopillars (Fig.~\ref{fig:structure}b)
as well as a 7th-stage Fibonacci structure consisting of $22$ nanopillars (Fig.~\ref{fig:structure}c). The
pillars are arranged according to 
Eqs.~(\ref{eq:cantor_analytic})~and~(\ref{eq:fib_analytic}), respectively.
Three-row Cantor and Fibonacci structures were made by putting 3~single-row 
structures in parallel at a distance~$\Delta$.
Fig.~\ref{fig:structure}d shows a 3-row Cantor structure for~$\Delta=0.75a$. 
As a reference system (also called {\em point-defect system} in what follows),
we consider a periodic array of the same nanopillars 
(spacing $d_L$) with a point defect in the middle\cite{Johnson} (Fig.~\ref{fig:structure}a). 
In analogy to 1D multilayer structures \cite{PRE}, we have
chosen a defect with the highest Q-factor possible, given a fixed 
number of pillars or layers. The defect is created by doubling the radius of
one of the nanopillars. This is
advantageous over, say, eliminating one pillar or reducing its radius,
because the defect mode is mostly localized within the optically dense medium,
thus reducing radiation losses.

The temporal evolution of fields in these systems was calculated using the 2D
finite difference time domain (FDTD) method\cite{FDTD}. The waveguide was placed inside 
a computation area, which is a $7a\times22a$ rectangle, 
with 16 mesh points per unit length $a$. 
In order to simulate radiation boundary conditions, 
perfectly matched layer (PML), which is designed to absorb 
all outgoing radiation without any reflection \cite{PML}, surrounded the rectangular 
computation area.

\begin{figure}[h]
\includegraphics[clip,width=0.5\columnwidth]{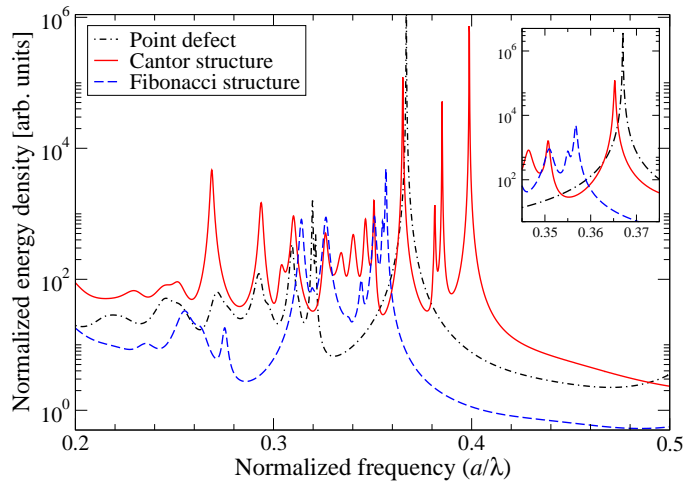}
\caption{Normalized energy density spectra of a Cantor (Fig.~\ref{fig:structure}b)
and Fibonacci (Fig.~\ref{fig:structure}c) waveguide versus a point-defect
reference waveguide (Fig.~\ref{fig:structure}a). The inset shows
the resonance peaks under study, located in the vicinity of the point
defect resonance. \label{fig:spectra}}
\end{figure}

To determine the spectral response of our systems, they were excited
by point dipole sources with random phase $\phi(\mathbf{r})$, so that at each point 
within the excitation area there is a current
\begin{equation}
\mathbf{j}(\mathbf{r},t)=-i\omega_{0}\mathbf{d}\delta(\mathbf{r})e^{-i(\omega_{0}t+\phi(\mathbf{r}))}e^{-\frac{(t-t_{0})^{2}}{\sigma}}.\label{eq:dipole}
\end{equation}
Evidently, this source is a Gaussian pulse (half width $\sqrt{\sigma}$)
of a dipole with base frequency $\omega_0$. The pulse was introduced
to make such sources emit a signal with a broad spectrum rather than
the $\delta$-function spectrum of a purely harmonic dipole source, where
$1/\sqrt{\sigma}$ was chosen large enough to cover the frequency window
of interest.
The spectral dependence of the energy density inside the system was then
obtained using Fourier transformation. The phase function~$\phi(\mathbf{r})$
in Eq.~(\ref{eq:dipole}) is taken to be random, and the calculation
was repeated with subsequent averaging over several
realizations. This is necessary in order to uniformly excite all possible
modes in the system. 

\begin{table}[b]

\caption{Normalized peak frequencies and corresponding Q-factors of some resonant
modes for single- and multiple-row nanopillar structures similar to
those shown in Fig.~\ref{fig:structure}.\label{table:qfactor}\\}

\begin{tabular}{|c|c||c|c|}
\hline 
Fig.&
Structure&
$a/\lambda$&
Q-factor\tabularnewline
\hline
\hline 
\ref{fig:structure}a&
Reference&
0.367&
$2.1\times10^{4}$\tabularnewline
\hline 
\ref{fig:structure}b&
Cantor &
0.365&
$2.7\times10^{3}$\tabularnewline
\hline 
\ref{fig:structure}c&
Fibonacci &
0.357&
$0.95\times10^{3}$\tabularnewline
\hline 
\ref{fig:structure}d&
3-row Cantor, $\Delta=0.75a$&
0.353&
$1.1\times10^{4}$\tabularnewline
\hline 
&
&
0.376&
$2.4\times10^{4}$\tabularnewline
\hline 
&
3-row Cantor, $\Delta=a$&
0.360&
$1.1\times10^{4}$\tabularnewline
\hline 
&
&
0.370&
$1.6\times10^{4}$\tabularnewline
\hline 
&
3-row Fibonacci, $\Delta=0.75a$&
0.343&
$1.1\times10^{3}$\tabularnewline
\hline 
&
&
0.356&
$1.1\times10^{3}$\tabularnewline
\hline 
&
&
0.366&
$1.6\times10^{3}$\tabularnewline
\hline 
&
&
0.373&
$1.7\times10^{3}$\tabularnewline
\hline
\end{tabular}
\end{table}

\subsection{Spectral response and resonant modes\label{sec:spectra}}

We now investigate the resonant modes in isolated, deterministically aperiodic
waveguides, i.e., without coupling to an external terminal, 
in order to justify that those resonant states can compete
with ``traditional'' defect states in terms of their Q-factor.
The spectra for Cantor and Fibonacci waveguides
vs.~reference system, normalized to the spectral response of free
space to eliminate the influence of the sources, are shown in Fig.~\ref{fig:spectra}. 
One can see that the reference system has a band gap above 
the normalized frequency $a/\lambda = 0.32$, where $\lambda$ is the
vacuum wavelength. In addition, the system exhibits a single sharp peak
inside the band gap at $a/\lambda\approx0.365$, which is obviously
associated with the defect. The Cantor waveguide does not exhibit a true 
band edge, but instead a crossover  
from the continuous spectrum (where the spectrum contains long-wavelength, effective-medium extended modes) to the fractal part of the spectrum (which primarily consists of resonant states intermingled with local band pseudogaps) \cite{fractons}. This crossover happens at a certain  
frequency (the phonon-fracton crossover frequency) found around 
$a/\lambda\approx 0.346$. The Cantor structure also exhibits a 
resonance peak very close to the resonance of the reference system. 
Since it is located above the phonon-fracton crossover, it is 
also associated with distortions with respect to a periodic lattice. For
the Fibonacci waveguide there is a group of peaks which form a transmission
band within a local pseudogap above $a/\lambda=0.3$. All structures in question
exhibit resonant modes at frequencies very near to each other, which makes these
modes suitable for comparative studies. Three-row waveguides tend
to have richer peak spectra resulting from the splitting (not shown here). 
However, several relatively isolated resonance peaks can usually be found
in the region of interest.

To estimate the resonant mode Q-factor, the system is excited
by a single point dipole oscillating harmonically at the resonant frequency.
After stationary oscillations were achieved, 
the dipole source was slowly turned off, and after that 
the subsequent field--time dependencies were analyzed in order to determine
the mode Q-factor as
\begin{equation}
Q=\frac{\omega\Delta t}{\alpha},\quad\alpha\equiv-\frac{d}{dj}\log\left\langle
\left|\mathbf{E}(j)\right|\right\rangle \label{eq:qfactor}\ ,
\end{equation}
where $\mathbf{E}$ is the electric field, $\langle \dots \rangle $ denotes the average 
over sufficiently many time steps $j$ so as to include a lot of light
oscillations, and $\Delta t$~is the time step width.
The results are compiled in Table~\ref{table:qfactor}. 

One can see that for a single-row Cantor waveguide the resonant mode
Q-factor is about one order of magnitude less than that for the point
defect. Various peaks within the above-mentioned Fibonacci pass band
generally have a Q-factor a couple of times less than that for Cantor
structure. As shown in Table~\ref{table:qfactor}, in the three-row Cantor waveguides the 
Q-factor is seen to rise to higher
values and becomes comparable to that of a point-defect reference
waveguide. It is not the case, however, in multi-row Fibonacci waveguides where
resonant modes exhibit only a moderate increase of the Q-factor.

\begin{figure}[h]
\includegraphics[width=0.5\columnwidth]{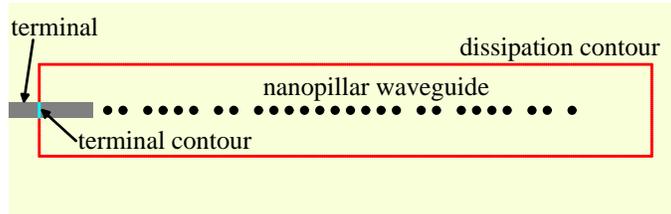}
\caption{The numerical set-up for energy flux calculations. The rectangular
closed contour is divided into a ``dissipation part'' outside the
terminal (\emph{red}) and the ``terminal part'' inside (\emph{cyan}).
The waveguide shown here is a Cantor structure with leftmost nanopillar
removed. The terminal width is $0.5a$.\label{fig:contour}}
\end{figure}

\section{Coupling to an external terminal\label{sec:Coupling}}

In the previous section we have shown that deterministic aperiodic
waveguides possess a Q-factor at least comparable to that of a point-defect
resonator. In this section, we demonstrate that Fibonacci and Cantor waveguides
can be more preferable in applications involving the 
energy exchange between the resonant system and the outside world in a
controlled way. By that we mean, that the balance between the ``useful radiation", i.e., radiation directed into the coaxial terminal, and the 
``lost radiation", i.e., radiation directed out of the waveguide axis, can be 
considerably shifted in favor of the ``useful radiation" in the case of proposed 
aperiodic structures without a dramatic compromise to Q-factor.

The \emph{terminal} introduced here is an unstructured dielectric rectangular
waveguide coaxial with the nanopillar waveguide and positioned at one of its ends. 
Since the terminal goes all the way into the PML boundary layer,
it can be considered semi-infinite from the point of view of our model.
To determine the spectral energy exchange characteristics of the system,
we place a single Gaussian-pulse point dipole source, as described by 
Eq.~(\ref{eq:dipole}), in the center of the structure and
analyze the Poynting vector $\mathbf{S}$ associated with the electromagnetic
fields $\mathbf{E}$ and $\mathbf{H}$ at frequency $\omega$, averaged over 
time, i.e.
\begin{eqnarray}
\overline{\mathbf{S}(\omega,\mathbf{r})}&=&
\int _0^{2\pi/\omega} \frac{\omega\, dt}{2\pi} \left| 
\mathbf{E}(\omega,\mathbf{r})\times\mathbf{H}(\omega,\mathbf{r}) 
\cos ^2(\omega t)\right| \nonumber \\
&=&\frac{1}{2}\textrm{Re}[\mathbf{E}(\omega,\mathbf{r})\times
\mathbf{H}^{*}(\omega,\mathbf{r})].\label{eq:poynting}
\end{eqnarray}
Since our computational scheme is effectively two-dimensional, 
the total energy flux at frequency $\omega$ 
is an integral over a contour enclosing the waveguide structure. 
If the contour contains no structure, the energy flux takes the form 
\begin{equation}
S_{0}(\omega)\equiv\oint\overline{\mathbf{S}_{0}(\omega,\mathbf{r})}\cdot\mathbf{n}dl\sim\frac{\sigma}{2}\omega\omega_{0}^{2}e^{-\frac{\sigma}{2}(\omega-\omega_{0})^{2}},\label{eq:poynting_closed}
\end{equation}
naturally independent of the closed contour shape. For computational simplicity,
it was chosen to be rectangular, as shown in Fig.~\ref{fig:contour}. 
Equation~\ref{eq:poynting_closed} has been confirmed 
by numerical calculations.

For a source radiating into the empty space the energy flux is isotropic, so the flux~$S_f$ through any contour fragment carries the same frequency dependence as in 
Eq.~\ref{eq:poynting_closed}, attenuated proportional to the aperture angle of 
that fragment~$\theta_f/2\pi$. 

As soon as a nanopillar structure supporting resonant modes is
placed inside the contour, with as well as without the terminal,
a strong redistribution of energy flux 
takes place. This redistribution is both spectral and spatial. 
The spectral one is seen as peaks at resonant frequencies,
where energy is stored within the structure
for a dwell time given by the inverse resonance width. 
The spatial redistribution accounts for the
pattern according to which the energy is radiated. 
In order to quantify the amount of energy directed into the terminal
compared to the amount of energy lost by radiation into the 
continuous modes of the surrounding homogenous medium (e.g. air),
we break our closed contour into a ``terminal part'', spanning the
cross-section of the terminal, and a ``dissipation part'' elsewhere
(see Fig.~\ref{fig:contour}).  The contour integrals of
$\overline{\mathbf{S}(\omega,\mathbf{r})}$
along these two fragments are named {\em terminal} and {\em dissipative} fluxes,
$S_{t}(\omega)$ and~$S_{d}(\omega)$, respectively.
The \emph{flux ratio} 
\begin{equation}
\eta(\omega)\equiv S_{t}(\omega)/S_{d}(\omega)
\end{equation} 
is then a quantitative measure of the balance between "usefull" 
and "lost" radiated power. For device applications involving a microresonator
mode coupled to an optical circuit both $\eta$ and $Q$ should be maximized. 
If we are to characterize the
resonant modes themselves with respect to spatial energy redistribution,
independent of the aperture of the terminal,
it is useful to compare the angular flux densities, i.e. the fluxes
$S_t$ and $S_d$ devided by their respective 
aperture angles $\theta_{t}$ and $\theta_{d}$, as seen from the
source position. In this way one obtains the 
\emph{flux density ratio} 
\begin{equation}
\zeta(\omega)\equiv S_{t}(\omega)\,\theta_{d}/S_{d}(\omega)\,\theta_{t}.
\end{equation} 
To eliminate the influence of the source spectrum on $S_t(\omega)$ and
$S_d(\omega)$, one can normalize the energy fluxes by~$S_{0}(\omega)$, 
Eq.~(\ref{eq:poynting_closed}). 

\begin{figure}[ht]
\includegraphics[clip,width=\columnwidth]{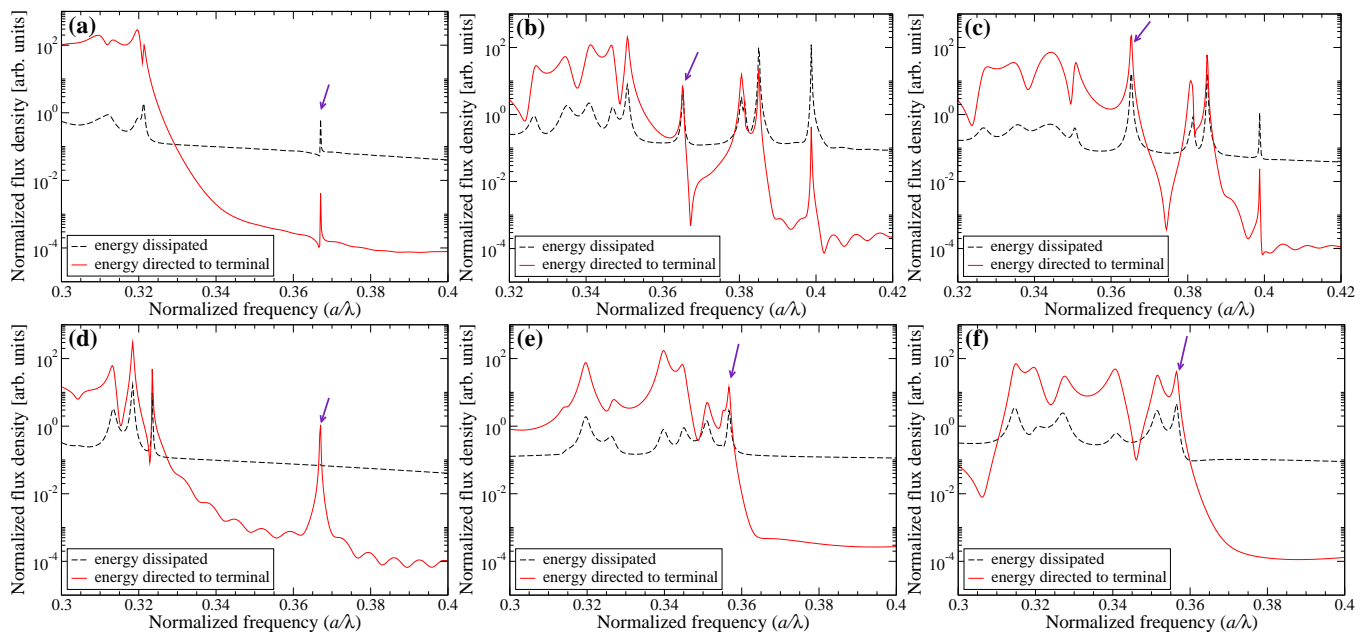}
\caption{The spectral energy flux densities, normalized to the total spectral flux
$S_0(\omega)$, through the terminal, $S_t(\omega)/\theta _t$, (red solid
line) and elsewhere, $S_d(\omega)/\theta _d$, (black dashed line) 
for the reference point-defect
structure \textbf{(a)}, a Cantor waveguide without \textbf{(b)} and
with \textbf{(c)} structure symmetry breaking, and a Fibonacci waveguide
without \textbf{(e)} and with \textbf{(f)} structure symmetry breaking.
The plot \textbf{(d)} corresponds to a point defect located 3 nanopillars
away from the terminal. \label{fig:poynting}}
\end{figure}

The results of numerical calculations are depicted in Fig.~\ref{fig:poynting}. We can see
at once that in a band gap the energy is mainly scattered out of the nanopillar waveguide, 
with~$S_{t}$ less than~$S_{d}$ by 4-5 orders of magnitude. This is the case
for all structures under study. Taking into account the ratio of aperture
angles, which in our setup equals~$0.00898$ (so roughly $\zeta\approx100\eta$),
one sees that $\zeta\simeq10^{-2}\ldots10^{-3}$. This is so because a structure
with a band gap cannot function as a
waveguide, prohibiting wave propagation in the direction of its axis
and thus radiating the energy mostly into the other directions.
On the other hand, in the pass band of a waveguide (especially in
the case of the reference system, Fig.~\ref{fig:poynting}a), the
energy directed into the terminal can approach or even exceed the
energy radiated elsewhere, so that $\eta\simeq1$ and $\zeta\simeq10^{2}$.
This is quite natural because the corresponding modes are extended and, thus, 
have a very good coupling to the terminal. However, it also means inevitably
a low Q-factor.  
Looking at the defect mode in the reference system (Fig.~\ref{fig:poynting}a),
we can see that although there is a peak in both energy fluxes, the
terminal flux is nearly as marginal compared to the dissipative flux
as in the gap ($\eta \approx 10^{-2}$). 
This means that, although this defect mode has a high
Q-factor, most of the energy is radiated in the transverse derections.

For the Cantor waveguides the situation
is different. As seen from Fig.~\ref{fig:poynting}b, 
for the resonant peak identified in Table~\ref{table:qfactor},
a considerable fraction of energy is directed into the terminal, the
ratio~$\zeta$ being around~$1$. Off-peak,~$\zeta$
decreases substantially, 
which identifies the peak as corresponding to a resonant mode. For 
the Fibonacci waveguide (Fig.~\ref{fig:poynting}e), 
the group of peaks in Fig.~\ref{fig:spectra} generally
resembles a pass-band bahavior, but with a lower~$\zeta\approx 1\, \dots\, 10$,
with the slight exception for the two rightmost peaks.
However, the Q-factor of these peaks is diminished (c.f. Table~\ref{table:qfactor})
due to the proximity to the pass
band and due to a more extended nature of the Fibonacci eigenstates. 

For both Cantor and Fibonacci waveguides, 
the coupling can be improved further, if the nanopillar
closest to the terminal is removed from the waveguide and the terminal 
is extended accordingly to keep the terminal--waveguide distance unchanged
(for the Cantor structure, compare Fig.~\ref{fig:structure}b 
with Fig.~\ref{fig:contour}).
In this case the internal symmetry of the structure, as implied by 
Eqs.~(\ref{eq:fib_example}) and (\ref{eq:cantor_example}), becomes distorted. 
Looking at Figs.~\ref{fig:poynting}c and \ref{fig:poynting}f
for Cantor and Fibonacci waveguides, respectively, one can see that for
the $a/\lambda = 0.365$ resonance,
$\zeta$ can amount on-peak to $10$ while off-peak it is still
far below~$1$. This means that a substantial part of resonant
mode energy is coupled into the terminal. Such a situation does not
occur for all peaks, however. For example, the low-Q, pass-band states for $a/\lambda \lesssim 0.36$ as well as the resonance at $a/\lambda \approx 0.4$ for Cantor structures retain their behavior as in the structure with unbroken symmetry.

\begin{figure}[th]
\includegraphics[width=\columnwidth]{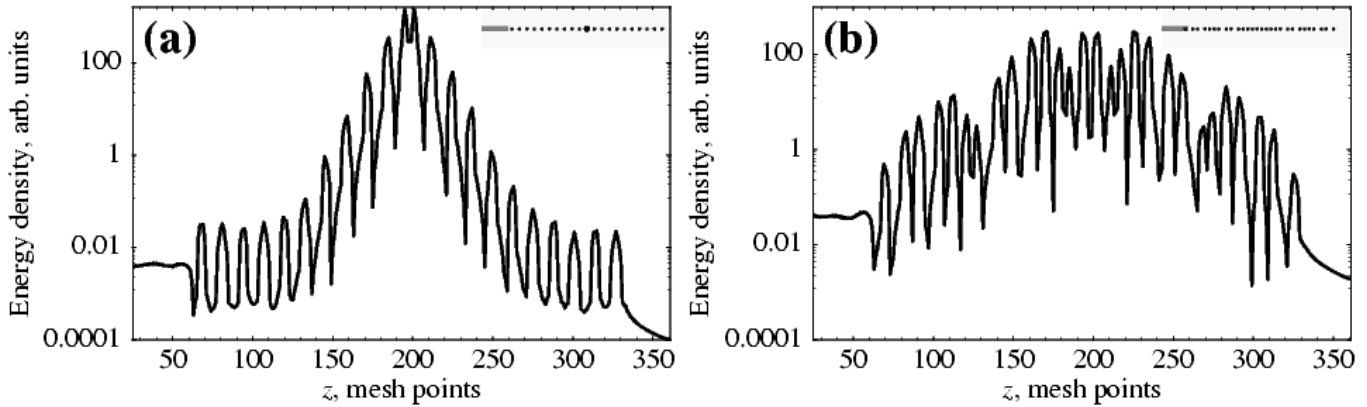}
\caption{The energy density profile along the longitudinal waveguide axis 
for the reference point-defect \textbf{(a)} and Cantor \textbf{(b)}
systems respectively. The corresponding structures are shown as insets.
The terminal is located to the left of each structure. \label{fig:decay}}
\end{figure}

Let us point out for comparison that altering the point-defect waveguide 
(Fig.~\ref{fig:structure}a) in a like manner, i.e., removing 
its leftmost pillar, does not change the coupling efficiency.
Only by moving the defect as close as~$3$ pillars away from the terminal
increases the coupling to comparable values with the Cantor structure case
(see Fig.~\ref{fig:poynting}d). However,
this symmetry breaking is accompanied by a drop in Q-factor 
by as much as two orders of magnitude, while removing one pillar 
from aperiodic (Cantor or Fibonacci) waveguides causes only a 10\% decrease.

\begin{figure}[th]
\includegraphics[clip,width=\columnwidth]{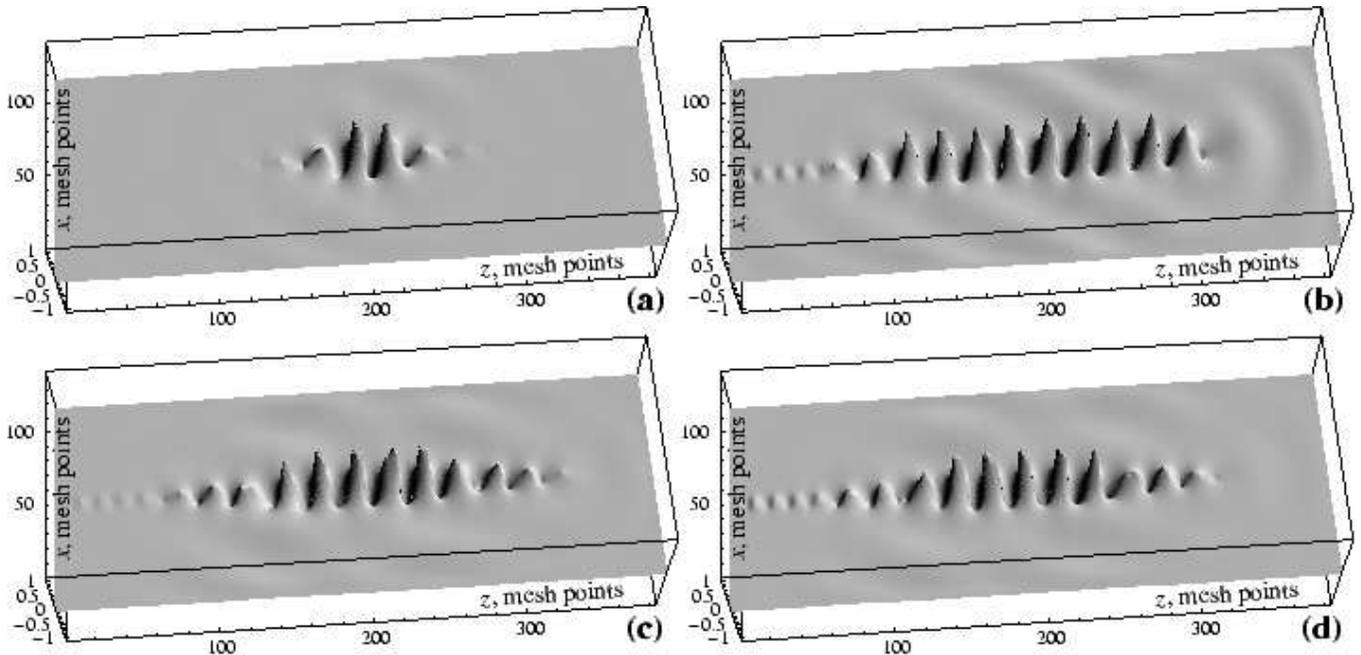}
\caption{Electric field distributions, normalized to their
respective maximum values, for the resonant modes under study
in presence of an external terminal: in the reference point defect
structure as seen in Fig.~\ref{fig:structure}a \textbf{(a)}; in
the Fibonacci \textbf{(b)} and Cantor \textbf{(c)} structure as depicted
in Fig.~\ref{fig:structure}b,c; in the same Cantor structure when
the terminal in positioned so as to cover the first nanopillar and
thus break the structure internal symmetry \textbf{(d)}. The system was excited
by a single monochromatic dipole source in the center of the structure
at the respective resonance frequency.
\label{fig:time-profiles}}
\end{figure}

Now we discuss the origin of the radiated power redistribution in favor of 
improved coupling to the terminal in the case of aperiodic structures as compared to periodic waveguides with a point defect. Figure~\ref{fig:decay} shows the energy density in a point defect structure (a) and a Cantor waveguide (b) along the axis of the structure,
when excited by a harmonic point defect in the center of the structure.
It is seen that in the periodic structure, the defect mode is strongly
exponentially localized around the point defect, and due to this exponential 
character the coupling is very poor. 
On the other hand, in the Cantor structure the energy density 
spreads out far more to the ends of the structure, while maintaining 
a relatively high Q-factor (c.f. Table \ref{table:qfactor}), leading to 
an enhanced coupling to the terminal and reduced radiation losses in 
the transverse direction. Note that this is not only due to the increased mode 
volume, but also because of a difference in mode localization character. 
We speculate that this is a precursor effect of the, on average, powerlaw localized, critical eigenmodes that are expected for an infinite aperiodic system, in analogy
to the critical eigenstates of quasicrystals \cite{Primer,Kohmoto}.

To visualize these results, we excited
the systems with a monochromatic point dipole 
at the respective resonant frequencies shown in 
Table~\ref{table:qfactor} and recorded the resulting electric
field distributions.
The results are shown in Fig.~\ref{fig:time-profiles}. In the point defect
(Fig.~\ref{fig:time-profiles}a), the field is strongly localized
with nearly isotropic radiation profile containing a nodal line along the waveguide. Hence,
nearly all the energy is dissipated sideways in accordance with
Ref. \cite{Johnson}. 
For the Cantor structures (Fig.~\ref{fig:time-profiles}c), 
there is a sizable field amplitude excited inside the terminal. 
The field in the surrounding area is comparable to that in the terminal, in agreement
with Fig.~\ref{fig:poynting}b. This behavior is further enhanced  
when the waveguide is used without its leftmost nanopillar,
i.e., if the terminal breaks the structure's 
symmetry (Fig.~\ref{fig:time-profiles}d).
The amplitude of sideways-directed radiation is the same as in Fig.~\ref{fig:time-profiles}c, 
but the field inside the terminal is greater, as implied by the condition $\zeta>1$.
Figure~\ref{fig:time-profiles}b shows the field distribution for the Fibonacci
structure. The resonant mode is less localized and the field inside 
the terminal is visibly more pronounced
than for the Cantor structure (Fig.~\ref{fig:time-profiles}c),
but so are sideways-directed radiation, resulting from smaller values of $\zeta$.

\section{Multiple-row Cantor waveguides\label{sec:Multiple}}

\begin{figure}[th]
\includegraphics[clip,width=\columnwidth]{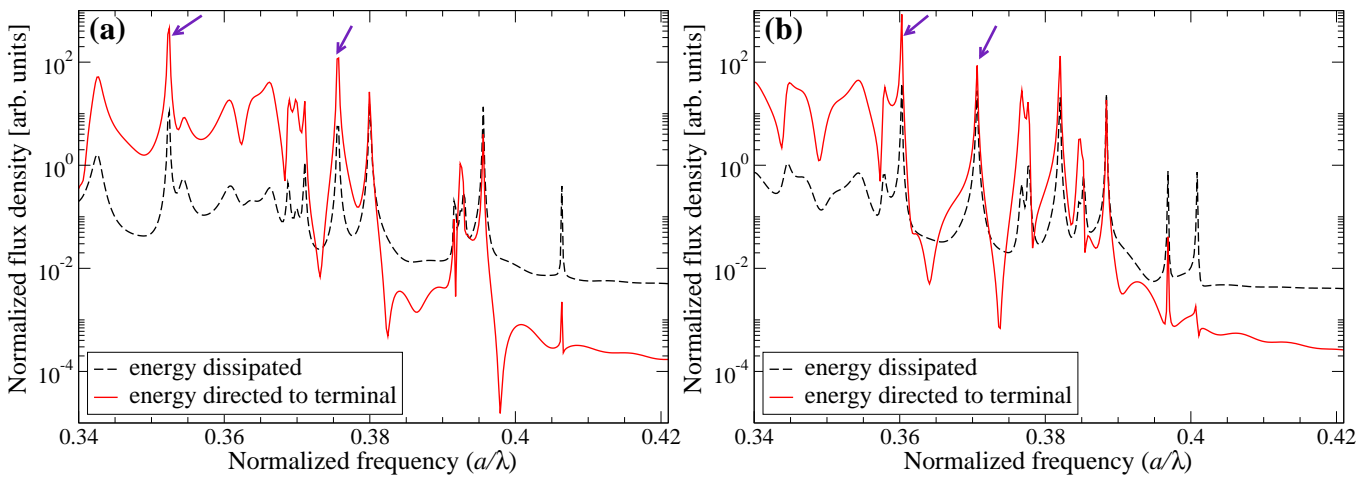}
\caption{The normalized energy flux through the terminal (red solid line)
and elsewhere (black dashed line) for a 3-row Cantor waveguide with
$\Delta=0.75a$ \textbf{(a)} and $\Delta=a$ \textbf{(b)}. Arrows
indicate the resonant peaks under study, 
shown in Table~\ref{table:qfactor}.\label{fig:multi-poynting}}
\end{figure}

\begin{figure}[th]
\includegraphics[width=\columnwidth]{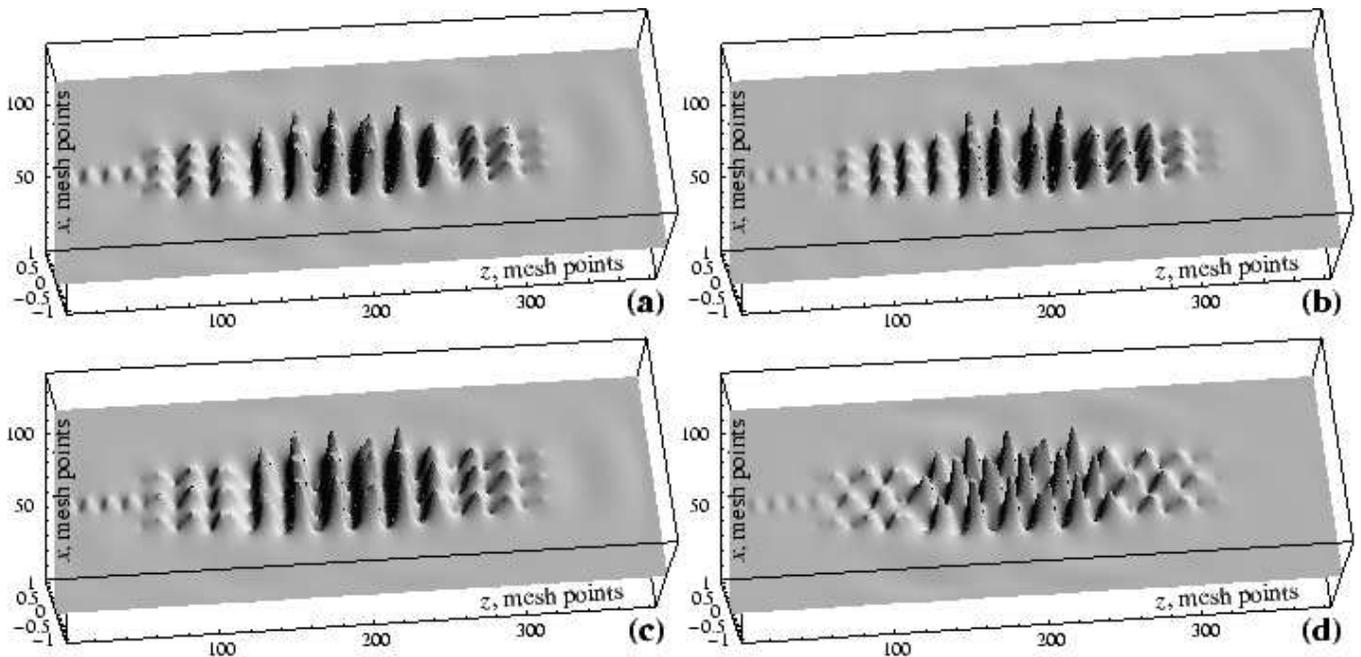}
\caption{Normalized electric field distribution of resonant modes for the
3-row Cantor nanopillar waveguide with the distance $\Delta=0.75a$
\textbf{(a,b)} and $\Delta=a$ \textbf{(c,d)} for the two different resonant
peaks (designated by arrows in Fig.~\ref{fig:multi-poynting}).
\label{fig:multi-time}}
\end{figure}

In Section~\ref{sec:Q-factor} we have shown that the resonant mode
Q-factors of Cantor nanopillar waveguides increase by almost a factor 
of ten and can compete with that of a point defect resonator, when 
arranged in a multiple-row fashion (see, e.g., Fig.~\ref{fig:structure}d).
In contrast, there is no substantial increase of Q-factor 
for Fibonacci structures. In this section, we therefore study the radiation pattern
of resonant modes in aperiodic multiple-row waveguides, 
resorting further to Cantor structures only.

As in the previous section, we begin the analysis by considering the
spatial redistribution of the energy fluxes into the terminal and
elsewhere for different frequencies in the spectrum. For the results to be comparable
to those obtained above, the terminal was chosen to be identical,
even though with respect to device design it might be advantageous
to increase its width along with
the lateral size of the multiple-row waveguide. According to our above
mentioned results, the terminal was positioned so as to break
the waveguide symmetry. Taking into account the width of the
terminal, this amounts to the leftmost nanopillar in the central
row being removed, and the terminal tip being inserted in its place. This
is obviously advantageous over removing the leftmost pillar from
all rows because the latter case would additionally increase radiation
losses.

The results are depicted in Fig.~\ref{fig:multi-poynting}. In general,
the picture looks much the same as in Fig.~\ref{fig:poynting}c, however
with an increased number of peaks and an increased Q-factor
of the modes (see Table~\ref{table:qfactor}). Detailed analysis
shows that for some peaks the ratio~$\zeta$ can be as high as 20--30,
providing an even better coupling between the mode and the terminal.
Thus, \emph{some resonant modes in multiple-row
Cantor waveguides combine a Q-factor as high as for the point defect
and an excellent coupling with the external terminal}. $\eta$ may be estimated 
to be 4--5 orders of magnitude better than in the reference point-defect
structure. 

For illustration, we have chosen two peaks for each value of
lateral spacing~$\Delta$, shown in Table~\ref{table:qfactor}.
The results for the electric field distribution are shown in 
Fig.~\ref{fig:multi-time}.
Comparing those with Fig.~\ref{fig:time-profiles}, we see that
the field amplitude inside the terminal is in general much greater
than that of the sideways radiation losses. 


\section{Conclusion\label{sec:Conclusion}}

To summarize, we have theoretically investigated resonant modes in
single- and multiple-row deterministically aperiodic nanopillar
waveguides, in comparison to a point defect in a periodic reference system.

Our calculations indicate that for certain resonance frequencies the Fibonacci and Cantor nanopillar waveguides do exhibit relatively high Q-factors along with improved coupling of the radiation to an external terminal. In gereral associated Q-factors are up to the order of magnitude lower with respect to the reference defect structure, while the amount of the radiation coupled to the terminal can be up to several orders of magnitude larger in the aperiodic structures. These may be traced back to the greater spatial extension of the modes along the aperiodic waveguide and to the strong anisotropy of their radiation pattern, 
with large fraction of  power radiated into the direction of the longitudinal
waveguide axis. The coupling to the external waveguide can be further enhanced
when the structure's internal symmetry is broken by the waveguide overlapping spatially with the aperiodic structure. Even a slight symmetry breaking (removing one nanopillar
from the terminal end of the structure) induces a significant 
coupling increase, while reducing the Q-factor by only about 10 per cent.
We found by contrast that, in order to achieve a similar coupling
in a periodic waveguide with point defect, 
one would need to place the defect as close as 3 nanopillars away 
from the terminal, which would, however, diminish the Q-factor by
two orders of magnitude.

The Q-factor of single-row aperiodic waveguides is in general smaller
by roughly an order of magnitude as compared to a point-defect resonator 
in a periodic structure. However, it becomes comparable to that 
of the reference point-defect system when multiple-row Cantor waveguides 
are used. Multiple-row Fibonacci waveguides do not show a significant
Q-factor increase.

We conclude that, in terms of simultaneous Q-factor and 
terminal coupling ($\zeta$) optimization, the Cantor structure 
appears to be the most favourable one among the ones studied here. However, we have only studied two out of many cases of deterministically aperiodic sequences. 
In this context it is worth to investigate other quasiperiodic and fractal
structures as well. Deterministically aperiodic 
structures can also be employed in the design of other types of photonic crystal waveguides, which is also a promising subject of further research. For example, by 
inserting 1D quasi-periodic sequence of rods or holes into a photonic 
crystal waveguide one can introduce new resonances into the 
transmission characteristic of the waveguide, while the scattering loss 
should be dramatically reduced due to the presence of the full 2D photonic 
bandgap in the surrounding 2D lattice.

\begin{acknowledgments}
The authors are grateful to A.~V.~Lavrinenko for taking a major
part in the FDTD/PML computation code design. This work was supported
by the Deutsche Forschungsgemeinschaft through SPP 1113 and 
Forschergruppe 557.
\end{acknowledgments}

\end{document}